# Two-Dimensional Short-Range Chemical Ordering in Ba$_{1-x}$Na$_x$Fe$_2$As$_2$

R. Stadel,[1,2*] R. DeRose,[1,2] K. M. Taddei,[3] M. J. Krogstad,[4] P. Upreti,[1,2] Z. Islam,[4] D. Phelan,[1] D. Y. Chung,[1] R. Osborn,[1] S. Rosenkranz[1] and O. Chmaissem[1,2]

[1]*Materials Science Division, Argonne National Laboratory, Lemont, IL 60439, USA*
[2]*Physics Department, Northern Illinois University, DeKalb, IL 60115, USA*
[3]*Neutron Scattering Division, Oak Ridge National Laboratory, Oak Ridge, TN 37831, USA*
[4]*Advanced Photon Source, Argonne National Laboratory, Lemont, IL 60439, USA*
[*]*Current Affiliation: Chemistry and Biochemistry, University of Maryland, College Park, MD 20742, USA*

## Abstract

A true understanding of the properties of pnictide superconductors require the development of high-quality materials and performing measurements designed to unravel their intrinsic properties and short-range nematic correlations which are often obscured by extrinsic effects such as poor crystallinity, inhomogeneity, domain formation and twinning. In this paper, we report the systematic growth of high-quality Na-substituted BaFe$_2$As$_2$ single crystals and their characterization using pulsed-magnetic fields x-ray diffraction and x-ray diffuse scattering. Analysis of the properties and compositions of the highest quality crystals show that their actual Na stoichiometry is about 50-60% of the nominal content and that the targeted production of crystals with specific compositions is accessible. We derived a reliable equation to estimate the Na stoichiometry based on the measured superconducting T$_C$ of these materials. Attempting to force spin reorientation and induce tetragonality, orthorhombic Ba$_{1-x}$Na$_x$Fe$_2$As$_2$ single crystals subjected to out-of-plane magnetic fields up to 31.4T are found to exhibit strong in-plane magnetic anisotropy demonstrated by the insufficiency of such high fields in manipulating the relative population of their twinned domains or in suppressing the orthorhombic order. Broad x-ray diffuse intensity rods observed at temperatures between 30 K and 300 K uncover short-range structural correlations. Local structure modeling together with 3D-ΔPDF mapping of real-space interatomic vectors show that the diffuse scattering arises from in-plane short-range chemical correlations of the Ba and Na atoms coupled with short-range atomic displacements within the same plane due to an effective size difference between the two atomic species.

**Introduction**

Since their discovery, hole-doped $A_{1-x}B_x$Fe$_2$As$_2$ (A = Alkaline Earth, B = Alkali metal) [1–6] and other "122"-type iron-based pnictides have played an important role in developing a comprehensive understanding of the magnetic order as an instability leading to unconventional electron pairing and induced superconductivity [7–12]. In this system, superconductivity [2,13–20] is achieved by suppression of an antiferromagnetic (AFM) stripe spin density wave (SSDW) ground state by the application of external pressures [21–23] or by charge doping via chemical [24] or isovalent substitution [25] much as in the cuprate superconductors. This natural tendency for superconductivity to arise from the suppression of magnetic order in a highly correlated electronic state with multiple fluctuating order parameters has been the basis for much of our current understanding of unconventional superconductivity and continues to inform the search for new superconductors.

At room temperature, the hole-doped "12*2*" materials crystallize in the tetragonal I4/*mmm* space group symmetry and exhibit paramagnetic properties. Compositions with low to intermediate substitution levels of the alkali metal (*i.e.*, x < 0.3-0.5 depending on the series being investigated), exhibit simultaneous first-order structural ($T_s$) and magnetic ($T_N$) phase transitions to the lower symmetry of the orthorhombic F*mmm* space group which also hosts the AFM SSDW. The phase diagrams of several "122"-systems with *A* = Ba, Sr, or Ca and B = Na, or K all exhibit similar features in which a narrow low temperature dome is identified deep within the SSDW region where the magnetic order changes from in-plane single-**Q** SSDW to an out-of-plane double-**Q** charge spin density wave state (CSDW) which coexists microscopically with superconductivity below $T_c$ [4–6,26–30]. The CSDW state causes the structure to return to its high temperature tetragonal symmetry. Due to its four-fold rotational symmetry, the re-entrant CSDW phase has been commonly known as the "$C_4$" phase. However, an in-plane four-fold rotationally symmetric magnetic double-**Q** Spin Vortex Crystal (SVC) state has also been observed in (*A,B*)-site ordered CaKFe$_4$As$_4$ [31,32] ("1144"), and more recently in LaFeAs$_{1-x}$P$_x$O (1111) [33], thus, necessitating the use of unambiguous notations such as $C_{2M}^{a}$, $C_{4M}^{ab}$ and $C_{4M}^{c}$ for orthorhombic in-plane single-**Q** SSDW, tetragonal in-plane double-**Q** SVC, and tetragonal out-of-plane double-**Q** CSDW, respectively. The subscripts indicate the two-fold orthorhombic (2M) or the four-fold tetragonal (4M) symmetry of the magnetic phase while the superscripts indicate the direction of the magnetic moment with respect to the nuclear/magnetic unit cell. Similarly, superscripts in the AFM Néel transition temperature notations, $T_N^a$ and $T_N^c$, will hereafter be used to denote the direction of the magnetic moment along the in-plane *a*-axis or out-of-plane *c*-axis, respectively.

The orthorhombic distortion in the 122 systems is rather small, in particular in the region where the structural and magnetic transitions are suppressed, and superconductivity is induced. The tendency to form microscopic or submicroscopic twin domains in such structures with almost identical lattice parameters has adverse consequences on our understanding of the intrinsic electronic and magnetic anisotropies that are believed to be of great importance in the iron pnictides as is revealed by a great deal of theoretical [34–37] and experimental [38–49] work. In powder diffraction studies, such small distortions due to nematic fluctuations can be resolved by pair distribution function (PDF) methods, where the relevant interatomic distances between neighboring atoms are analyzed over variable length-scales.

The manipulation of the relative population of the orthorhombic domains and the realization of nearly 100% twin-free single crystals was successfully achieved with uniaxial pressures [48,50] or strong magnetic fields applied in the in-plane direction [51,52]. Detwinned crystals enabled investigations of the intrinsic magnetic order properties of electron-doped Ba(Fe$_{1-x}$Co$_x$)$_2$As$_2$ single crystals [36,38], for example, which revealed the existence of large in-plane anisotropy in addition to nematic ferro-orbital ordering at temperatures much higher than $T_N$ and $T_s$. However, what hasn't been examined is whether or not the application of a strong magnetic field along the *c*-axis could coerce the SSDW in-plane moments to rotate in the direction of the out-of-plane magnetic field, particularly when applied at temperatures near the structural and $C_{4M}^c$ magnetic transition. A forced out-of-plane alignment of the Fe magnetic moments may potentially suppress the material's orthorhombicity and induce tetragonality.

In addition to twinning, effects of chemical substitutions could also obscure investigations of intrinsic nematic anisotropies. In particular, whenever a crystallographic site is shared by two or more atomic species, several outcomes are possible depending on the ionic radii and chemical affinity of the selected elements. In the Ba$_{1-x}$Na$_x$Fe$_2$As$_2$ system, the Ba and Na atoms could be randomly distributed on the shared site, locally ordered in distinct clusters with various Ba/Na ratios and bonding configurations, or fully long-range ordered where the atoms would become independent and occupy different crystallographic sites. The first and second situations can be easily distinguished experimentally by the shape of the diffuse scattering, whereas the latter case is achieved in the fully-ordered "1144" line compositions, for example. To better understand the stability and formation of the "1144" structures, Song et al. [53] performed density functional theory calculations of the enthalpy for a large number of materials and identified a significant number of compositions that favor formation of the "1144" structure. Furthermore, the authors determined that even some well-known "122" materials like Ba$_{0.5}$Rb$_{0.5}$Fe$_2$As$_2$ and Ba$_{0.5}$Cs$_{0.5}$Fe$_2$As$_2$ energetically favor the "1144" instead, thus, suggesting that *A* and *B* atomic ordering may be possible by prolonged annealing of the materials. Interestingly, in a prior study, Iyo et al. [54] established the existence of a linear relationship between the differences in lattice parameters, $\Delta a = (a^{A122} - a^{B122})$, and ionic radii, $\Delta r = (r^A - r^B)$, in which the "1144" and "122" phases are clearly separated with Ba$_{0.5}$Rb$_{0.5}$Fe$_2$As$_2$ and Ba$_{0.5}$Cs$_{0.5}$Fe$_2$As$_2$ being right in the middle between the two regimes. Ba$_{0.5}$Na$_{0.5}$Fe$_2$As$_2$, on the other hand, is found at the far end of the 122 region, thus indicating that it is highly unlikely to form any long-range ordered 1144-type structure regardless of the synthesis conditions. While this is technically true, we will demonstrate that short-range correlations of the Ba and Na atoms do occur.

In this work, experiments were performed to address key questions relative to the strength of in-plane vs out-plane magnetic anisotropy for Ba$_{1-x}$Na$_x$Fe$_2$As$_2$ compositions near the $C_{4M}^c$ dome, and to the nature of intrinsic short-range nematic correlations often-entangled with extrinsic effects such as domain twinning or sample inhomogeneity. We report the growth of high-quality Ba$_{1-x}$Na$_x$Fe$_2$As$_2$ single crystals and a method for the quick evaluation of their stoichiometry based on easily performed measurements of $T_c$. Using well characterized crystals, high-resolution x-ray diffraction in pulsed magnetic fields up to 31.4T along the out-of-plane direction was used to demonstrate the strong in-plane magnetic anisotropy of Ba$_{1-x}$Na$_x$Fe$_2$As$_2$ and the inability of such high fields in de-twinning the crystals or in driving spin reorientation of the Fe magnetic moments. Further, our diffraction data bring to light the presence of anisotropic local structures in the form of diffuse intensity rods and blobs that we successfully model as two-dimensional short-range chemical ordering of the Ba and Na atoms in the *ab* plane with no correlations observed along the *c*-axis. This ordering, albeit over short length-scales, is different than the layer-ordering of the *A* and *B* ions in the 1144 analogs. Together, these results show a surprising stability of the $C_{2M}^a$ magnetic structure and that local correlations due to chemical substitution can mimic intrinsic nematic correlations.

**Experimental Details**

With focus dedicated to compositions within or near the magnetic $C_{4M}^c$ dome, systematic single crystal growth batches of targeted $Ba_{1-x}Na_xFe_2As_2$ compositions were performed in multiples with the nominal Na stoichiometry x = 0.30, 0.33, 0.34, and 0.35. Additional batches were prepared with higher Na contents x = 0.45, and 0.5. Single crystals were grown using elemental Na and binary FeAs, $Fe_2As$ and "BaAs" precursors with excess FeAs acting as self-flux. Experiments with relatively similar conditions have been reported in the literature for the growth of $Ba_{1-x}Na_xFe_2As_2$ and $Sr_{1-x}Na_xFe_2As_2$ single crystals [46,55–57]. All the chemicals were handled in a glovebox filled with argon with <5 ppm oxygen and <0.1 ppm $H_2O$. The stoichiometric binary precursors were pre-synthesized in evacuated quartz tubes, however, because of the strong tendency of Ba to react with quartz at high temperatures, raw mixtures of BaAs powders were loaded in cylindrical alumina crucibles before encapsulation in the tubes. Each quartz tube was initially annealed at 700°C for 24-48 hours after which it was opened, the precursor thoroughly homogenized, and then repacked and-annealed using the same procedures. The quality of the precursors was verified by x-ray diffraction prior to crystal growth with further annealing cycles performed as necessary.

Due to the limited volume of the sealed quartz tube and to avoid excessive pressures that may lead to tube rupture, crystal growth in batches of no more than 5-7 grams was performed by placing the volatile elemental Na at the bottom of an alumina crucible over which well-mixed precursor $Ba_{1-x}Na_xFe_2As_2$ powders and the FeAs flux were added in a molar ratio of 1:2. The alumina crucible was loaded inside a Nb ampoule, which was arc-welded under 1 atmosphere of argon, then loaded and sealed in an evacuated quartz tube. The tube was placed vertically in the furnace to prevent liquid flux from creeping out. The heating profile culminated in holding the tube at 1100°C for 12 hours before cooling to 750°C at a rate of 2°C/hour at which point the furnace was shut off and the samples allowed to cool naturally to room temperature. This growth procedure produced a cylindrically shaped boule containing all the reaction components with roughly three regions of varying quality and stoichiometry. The upper third of the boule resembled globs of brittle metals where the extra flux and Na seem to have accumulated. The bottom two thirds contained a large number of crystals that can be extracted by breaking the boule which causes it to cleave across the entire diameter along its vertical growth axis. Relatively thin crystals up to several millimeters in length and width were separated, examples of which can be seen in Fig. 1(a-d). Each boule produced many single crystals with the highest quality crystals, growing in the bottom third of the boule, tending to be ~1×1×0.05 mm$^3$ in size or smaller. Single crystals extracted from the middle third had higher $T_C$s, indicative of higher Na contents, but with significantly wider superconducting transitions. Representative samples from each batch were crushed for x-ray powder diffraction, which showed patterns consistent with very pure $Ba_{1-x}Na_xFe_2As_2$.

      Superconductivity was determined using a Quantum Design Magnetic Property Measurement System (MPMS) with zero field cooling and a measuring field of 200 Oe on warming, Fig. 1e. X-ray scattering was performed on several crystals using a STOE x-ray diffractometer and at the beamlines 6-ID-C and 6-ID-D of the Advanced Photon Source (APS) at Argonne National Laboratory. Diffraction experiments at beamline 6-ID-C were performed in transmission mode at a beam energy of 20 KeV in zero magnetic field and in pulsed magnetic fields (PFXRD) up to 31.4 T at various temperatures between 16 K and 300 K [58]. For this experiment, the crystals were glued to a sapphire plate using GE varnish as shown in Fig. 2, and the plate glued to a sapphire rod was mounted at the center of a pulsed magnet and held at temperatures just above and below $T_N^a$ and $T_N^c$ and at 16 K. Fig. 3 shows an example of the instrument's oscilloscope readout recorded during magnetic pulsing. Several frames were taken prior to the pulse with the pulse peak approximately centered on the 17th frame while the

remaining frames captured the structural response of the crystal over the last ~27.5ms. Data were collected using a two-dimensional MMPAD detector placed approximately 2.4 m from the sample. The detector has 2 x 3 modules with 128 x 128 pixels each. Each pixel size is 150 $\mu$m x 150 $\mu$m [59,60].

In the 6-ID-D diffuse scattering experiment, a single crystal is typically mounted on a two-axis goniometer and placed at a distance of 650 mm from a Dectris Pilatus 2M area detector where it is exposed to a high-flux x-ray beam of wavelength ~ 0.142 Å (87.1 keV). A vertically mounted N-Helix cryostat was used to vary the sample's temperature between 30 K and 300 K.

Modeling of our diffuse scattering data was performed using DISCUS simulations and 3D-ΔPDF real-space mapping. A model single crystal containing 100 x 100 x 20 unit cells was created in DISCUS. All atoms other than Ba and Na were ignored. Ba and Na atoms, in the same nominal ratio, were initially distributed randomly and then allowed to order by a Monte Carlo (MC) simulation process to achieve nearest-neighbor correlations with Warren-Cowley [61] parameters $\alpha_{100}^{Ba-Na} \approx -0.083$ and $\alpha_{110}^{Ba-Na} \approx 0.064$, in rough agreement with the observed broad scattering. The Ba and Na atoms were then displaced via another MC simulation process, using Lennard-Jones potentials to increase the Ba-Ba distances and decrease those of the Na-Na type for neighbors at [100] and [210] interatomic vectors. Guided by our 3D-ΔPDF measurements, for the [110] neighbor, the Na-Na distances were increased while decreasing the Ba-Ba distances.

**Results and Discussion**

***Crystals stoichiometry, and superconducting and crystalline quality***

Each crystalline boule, grown under the conditions described above, allowed the extraction of many platelet-like Ba$_{1-x}$Na$_x$Fe$_2$As$_2$ single crystals of various qualities and stoichiometries. Because of the compositional gradient of volatile Na along the boule's vertical axis, the cleaved single crystals are expected to be stoichiometrically diverse. Therefore, prescreening the crystals for further measurements is necessary. Energy dispersive x-ray (EDX) spectroscopy and in-house single crystal x-ray diffraction techniques are generally good for investigating the crystal's structural and compositional properties; however, they are quite slow and sub-optimal at identifying homogeneity for crystals larger than the beam. Bulk Meissner measurements, on the other hand, can be quickly preformed and have the advantage of being very sensitive to subtle stoichiometric variations in the full crystal due to the significant dependence of the $T_c$ on the composition. In this work, the primary method we employed for determining the stoichiometry and quality of each crystal was measuring and analyzing the sharpness of its superconducting transition.

Measurements of the superconducting transition temperature $T_c$ (defined as the intersection of lines tangent to zero and to the steepest descent of the moment signal) of many crystals selected from each batch show sharp superconducting transitions as displayed in Fig. 1e. The superconducting transition width $\Delta T_c$, defined as the difference between $T_c$ and the temperature at which magnetic susceptibility reaches 90% of its final value, was used to characterize the sharpness of the crystal's superconducting transition. Crystals with broader transitions (excluded from this study) tended to be large and not fully homogeneous.

To determine the exact stoichiometry of the single crystal, the measured $T_c$ was mapped onto our published Ba$_{1-x}$Na$_x$Fe$_2$As$_2$ phase diagram [3,4]. Given the large number of high-quality samples used for its construction, we use this phase diagram to fit its relevant superconducting data points with the following cubic function:

$$T_C = 318.99x^3 - 683.89x^2 + 425.20x - 47.78 \quad Equation\ 1$$

with the equation being valid for superconducting compositions in the range $0.15 \lessapprox x \lessapprox 1$, Fig. 4. As expected from this function, it can be easily shown that small variations in the Na content (x) correspond to relatively large changes in $T_c$ in a small region of the phase diagram where the competing magnetic and superconducting states coexist. Equation 1 does a remarkable job in estimating the crystal's sodium content within narrow margins given that the highest quality crystals have $\Delta T_c$ of ~2-3 K (corresponding to $\Delta x \approx 0.014 - 0.02$). Here, it should be emphasized that equation 1 works well even for samples in the $C_{4M}^C$ dome, where superconductivity and magnetism coexist, because the slight suppression of $T_C$ [62] due to the tetragonal $C_{4M}^C$ phase in the Ba$_{1-x}$Na$_x$Fe$_2$As$_2$ system is roughly of the same magnitude as the magnetization measurements uncertainties. That is to say that the variance within a top-quality crystal is greater than the disagreement between the predicted and measured/refined stoichiometry of a given crystal (via x-ray diffraction for example). The nominal and fitted Na stoichiometries for several high-quality crystals are presented in Fig. 1e together with their normalized superconducting transitions.

With the highest quality crystals extracted from the bottom third of the boule, it is to be expected that their actual Na stoichiometry is significantly lower than nominal as confirmed by the systematic shift displayed in Fig. 5. Although our attempted batches include a significant gap between 0.35 and 0.45 in the nominal Na starting composition, we note that the actual Na content of the best crystals is approximately 50-60% of the nominal concentration and that very few crystals are obtained in the region where the $C_{4M}^c$ phase forms (*i.e.*, in the 0.22 ≤ x ≤ 0.30 range [3,6,62] – see samples within the $C_{4M}^c$ dome in the inset of Fig. 4). This region of relative metastability agrees well with experience gained from powder synthesis, from which we learned that achieving sample purity in this compositional range requires up to 8 or 9 cycles of homogenization and annealing compared to samples with more or less Na (for which 3 to 4 cycles are usually sufficient). It is worth emphasizing that even the batch with the lowest nominal Na (i.e., x = 0.3) contains more Na than needed for the formation of the delicate $C_{4M}^c$ phase, however, only the nominal x = 0.45 and x = 0.50 batches produced crystals measured within the $C_{4M}^c$ dome. Our results and the linear relationship between nominal and fitted Na stoichiometry, Fig. 5, suggest that targeting the $C_{4M}^c$ phase or any other desired composition requires twice as much excess Na in the starting batch mixture.

The robustness of equation 1 in determining the actual stoichiometry was tested by x-ray diffraction and magnetic measurements. The analysis of a crystal with nominal sodium content x = 0.35 and a superconducting transition temperature $T_c$ of 27.1K ($\Delta T_c$ ~ 5 K), for example, suggests that its Na stoichiometry (x) should be ~ 0.30(2). Indeed, single crystal x-ray diffraction and refinements performed using GSAS-II [63] with a weighted residual agreement factor (wR$_I$) of 4.5%, resulted in a refined Na content x = 0.292(1). Every crystal whose composition was measured by x-ray diffraction at low temperature had its orthorhombic structural transition confirmed at the temperature predicted from its fitted sodium content.

*X-ray Scattering*

In previous work, Ruff *et al.* [51] demonstrated the successful detwinning of a single crystal of underdoped Ba(Fe$_{1-x}$Co$_x$)$_2$As$_2$ using a 27.5T pulsed magnetic field applied in the in-plane direction at temperatures very close below and above its orthorhombic to tetragonal phase transition. Recent work [48], on the other hand, gave evidence that moderate uniaxial pressures between 20 and 45 MPa can be used to de-twin an underdoped BaFe$_2$As$_2$ single crystal when applied along the in-plane *b*-axis and that the crystal's magnetic easy axis rotates towards the *c*-axis by about 28° from the in-plane *a*-axis direction but only when performed at temperatures just below the nuclear and magnetic transition temperatures, $T_N/T_S$. This *c*-ordered magnetic moment component, observed only in uniaxially strained orthorhombic BaFe$_2$As$_2$, is markedly different than the out-of-plane tetragonal CSDW phase and suggests the existence of strong spin-orbital coupling responsible for induced magnetic anisotropy.

Two crystals with fitted Na content x = 0.17 and 0.21, close to the $C^c_{4M}$ dome, were selected for x-ray measurements under high magnetic fields. Measurements in zero magnetic fields were performed for the x = 0.21 single crystal. Pulsed magnetic fields up to 31.4 T were applied along the *c*-axis direction at temperatures close to $T_N$ and at 16 K in an attempt to force an out-of-plane rotation of the magnetic moments so that the orthorhombic distortion is suppressed in favor of the tetragonal magnetic CSDW phase.

Our instrumental setup provides sufficient resolution to resolve the expected long-range orthorhombic splitting as seen by the evolution of the tetragonal (220)$_T$ peak which splits, upon cooling, into four (200)$_o$/(020)$_o$ type peaks, Fig. 6. The measured $T_S$ of ~ 108 K confirms the fitted stoichiometry of this x = 0.21 crystal. Measurements taken below $T_S$ at magnetic fields less than 17.5 T showed no perceptible indication of the field influencing the structure, as shown in Fig. 7 (see panels (*d, e*)). At higher fields and over a wide temperature range, there was no noticeable effect from the magnetic pulse, with the exception of measurements in 31.4T fields at 95K, Fig. 7(*a-c, f*), just below its orthorhombic transition temperature, in which we observe a field-induced shift in the shape of the twin Bragg peaks. This measurement was repeated 10 times confirming reproducibility of the observed magnetoelastic response. We expect a Bragg peak to possibly shift both off-scattering plane as well as in scattering angle due to magnetostriction (MS) effects, if present. In this case, however, the Bragg peak's shift seems dominated by a tilt away from the scattering plane in the same direction for both twin partners. A pair of Bragg reflections that are conjugates due to the splitting of a single tetragonal peak is expected to either merge or move farther apart with any modification of the crystal symmetry. Since no obvious MS is observed for the other composition (x = 0.17), MS if present in this sample is too small to be resolved in the presence of a large field-induced tilt. We speculate that this sample may have been vulnerable to experiencing a torque due either to possible iron inclusions or to canting of some domains such that their $ab$-plane was partially parallel to the applied field, as the $C_{2M}^{ab}$ phase has an SDW-type AFM order, being ferromagnetically aligned along the *b*-axis. This ferromagnetic alignment along one axis is the mechanism by which a magnetic field can de-twin these crystals when applied along the $ab$-plane [51]. In this work, we saw no response from the magnetic pulse in either suppressing the orthorhombic order or breaking the tetragonal symmetry when applied below or above $T_N$, respectively, or in affecting the relative population of the twinned domains. Our results, albeit neutral, demonstrate the strong anisotropy of the SDW magnetic order despite its exposure to such high magnetic fields along the *c*-axis. Capturing the suspected torque effect is useful in demonstrating what the result from a magnetic pulse should look like and verifying that our setup was adequate to detect such effects.

Fig. 8a shows the results of diffuse scattering measurements of a Ba$_{0.74}$Na$_{0.26}$Fe$_2$As$_2$ crystal ($T_c$ = 22 K) in which broad diffuse features are observed in the $[h, k, 0]$ plane. In Fig. 8b, on the other hand, these features take the form of broad intensity rods of diffuse scattering parallel to *l*. Some extra thermal diffuse scattering can be seen close to integer values of *l,* but aside from this, the rods have no intensity modulation, indicating the underlying structure has no correlation along *c.* These rods do not have a constant position in the *hk* section of the Brillouin zone; as displayed in Fig. 8c, they are near ½½0 for smaller values of |**Q**| but shift progressively closer to the zone center as |**Q**| increases.

To understand the origin of these diffuse scattering intensity rods, real-space interatomic vector maps were produced by Fourier transform of the diffuse scattering intensities according to procedures described elsewhere [64,65]. The 3D-ΔPDF map displayed in Fig. 9a, shows positive (red) and negative (blue) short-range correlations, within the *ab* plane, of the real space interatomic vectors that connect neighboring barium and sodium atomic pairs. The positive and negative signals at a position indicate whether the interatomic vector is more probable (red) or not (blue) with respect to the average long-range structure, respectively, weighted by the atomic form factors. The rapid decay of the correlations' strengths a few unit cells away demonstrates the short range of the ordering between Ba and Na atoms. These correlations have a small net signal at nearest- and next-nearest-neighboring positions but mostly have a dipole nature. This indicates that the diffuse scattering is produced by a combination of two effects [64]: the small net signal at each lattice vector is proportional to the chemical short-range order component of barium and sodium, and the dipole signature indicates an atomic size effect. The chemical order component shows that Ba and Na atoms slightly tend to alternate within the *ab* plane, with Na atoms rarely being nearest neighbors in this plane (expressed via Warren-Cowley parameters, $\alpha_{100}^{Ba-Na} \approx -0.08$; $\alpha_{110}^{Ba-Na} \approx 0.06$); this short-range order will produce scattering near **q** = ½½0 and manifests as the small net signal in the 3D-ΔPDF. The atomic size effect component arises from the smaller effective size of Na relative to that of the Ba atoms, with the structure locally relaxing towards the Na atoms and away from Ba; this is seen in the 3D-ΔPDF as a negative signal on the low-|r| side from lower-*Z* Na-Na pairs and a positive signal on the high-|r| side from the larger-*Z* Ba-Ba pairs. This relaxation produces scattering on the low-|**q**| side of the zone center. The zone-boundary chemical order scattering decays with increasing |**Q**| more quickly than the zone-centered size effect scattering, which accounts for the shift toward the zone center with increasing |**Q**|. These two effects were investigated further via a simulation in DISCUS [66], Fig. 9b. This model only includes the Ba-Na sublattice, initially introducing chemical ordering via a Monte Carlo simulation process and then adding the size effect via a second MC process. Calculated x-ray scattering from this model before and after consideration of the size effect displacements is shown in Figs. 9(c-f), respectively. It is clear that the combination of these two effects qualitatively reproduce the broad scattering seen in the experiment. Taking a 2D-PDF cut from the calculated data provides some important insights: while the dipole-like size effect signatures are reproduced, there are quadrupole-like signatures in the experimental data (note particularly the 200 and 220 lattice vectors) that are not present in the simulation. This indicates the presence of displacement correlation in the observed data, not present in the model, possibly consistent with local orthorhombic fluctuations in agreement with previous PDF studies [67,68].

The rod-like nature of the diffuse scattering emphasizes that these short-range effects do not extend out of the $ab$ plane, with chemical occupancy being uncorrelated along the $c$-axis. The lack of chemical ordering along the $c$-axis has implications relating to the stability of the $A_1B_1Fe_4As_4$ phase (1144s) [54,69–72]. The 1144s are well-established as being a special case of 122s with ordered doubling of the $c$-axis due to alternating $A$ and $B$ cation layers [53,54]. Theoretical and experimental examinations [53,54] have been performed to determine under what circumstances and with what combination of cations this 1144 phase is stable. As mentioned earlier, Song *et al.* [53] showed that the sodium-doped 122s will not form a stable long-range 1144 structure, with the $Ba_{1-x}Na_xFe_2As_2$ being the least optimal hole-doped 122 system for the 1144 phase to form due to the difference in ionic sizes and Fe-As bond lengths (see figure 5 in reference [54]). In the light of the tendency of $Ba_{1-x}Na_xFe_2As_2$ to not form the 1144 structure, the lack of long-range correlation along the $c$-axis is to be expected. It is, however, surprising, and interesting to see no evidence of such correlations along $c$ on shorter length scales. We speculate that similar measurements on other Na-substituted 122 compositions and stoichiometries should reveal diffuse features in proportion to the ratio of ionic sizes and inverse proportion to Fe-As bond length, as those were found to be the deciding parameters in the formation stability of the 1144 structure in the aforementioned work [53].

In addition to the rods, there are recurring regular diffuse features (Fig. 8b) which present as nebulous intensities occurring along the $h$ plane on the $+|h|$ side of the rods (furthest from the origin) at integer values of $l$. The intensity of these features appears proportional to the structure factor of nearby Bragg peaks consistent with thermal diffuse scattering [73].

**Conclusions**

We have developed a reliable technique for the growth of high-quality Ba$_{1-x}$Na$_x$Fe$_2$As$_2$ single crystals and determined the existence of a systematic linear shift in their actual stoichiometry being roughly 50-60% that of the nominal Na content. Best nominal sodium conditions to grow crystals in the metastable $C_{4M}^c$ dome or other desired compositions are determined. A systematic method for verifying a crystal's stoichiometry with easily performed magnetic measurements was presented. X-ray scattering demonstrates a surprisingly strong in-plane magnetic anisotropy of the SSDW ($C_{2M}^a$) phase when subjected to high out-of-plane magnetic fields up to 31.4T. No detwinning or structural transformations are observed under such *c*-axis oriented magnetic fields. Zero-field three-dimensional x-ray measurements uncovered, on the other hand, the existence of broad diffuse scattering rods that reveal previously-unknown local chemical ordering in the *ab* plane with an effective size difference between barium and sodium atoms manifesting in local structural distortions and displacements. Our results emphasize the need for considering the local correlations effects and short-range site-ordering in investigations of the intrinsic nematic correlations in these materials. With the many fascinating features seen in this and related systems, a larger number of high-quality crystals with well-controlled doping is invaluable for future research.

**Acknowledgements**

This work was primarily supported by the US Department of Energy, Office of Science, Basic Energy Sciences, Materials Science and Engineering Division. This research used resources of the Advanced Photon Source, a U.S. Department of Energy (DOE) Office of Science User Facility, operated for the DOE Office of Science by Argonne National Laboratory under Contract No. DE-AC02-06CH11357. The high-field pulsed magnet and a choke coil were installed at the Advanced Photon Source through a partnership with International Collaboration Center at the Institute for Materials Research (ICC-IMR) and Global Institute for Materials Research Tohoku (GIMRT) at Tohoku University.

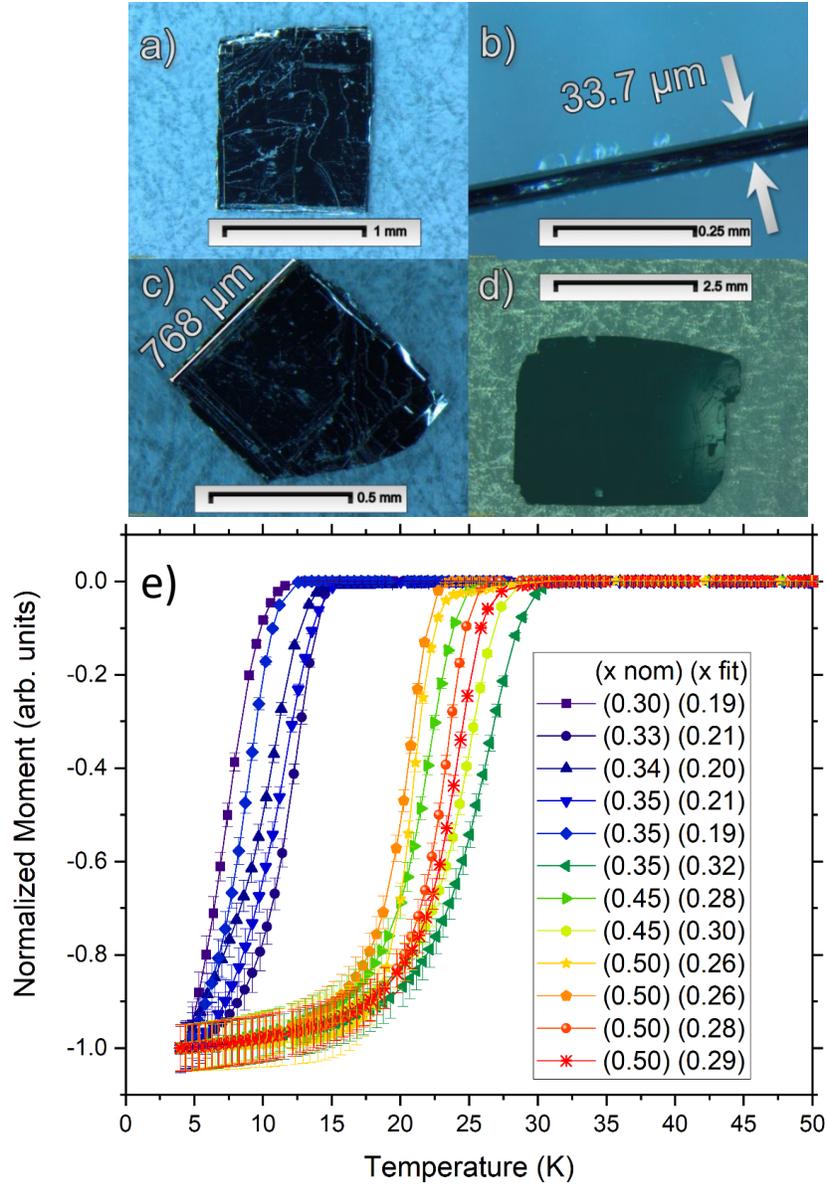

*Fig. 1 Single crystals extracted from three different batches (a, c, and d). a) and b) are top and side views of the same crystal. (e) Superconducting transitions of representative high-quality crystals extracted from the diverse batches. Nominal and fitted values for Na content (x) are shown in the inset legend. Raw magnetization signals (not shown) are quite different because of the varying sizes of the crystals being measured. Magnetization is normalized to -1 so that the $T_C$ and superconducting transition sharpness are emphasized rather than the signal itself.*

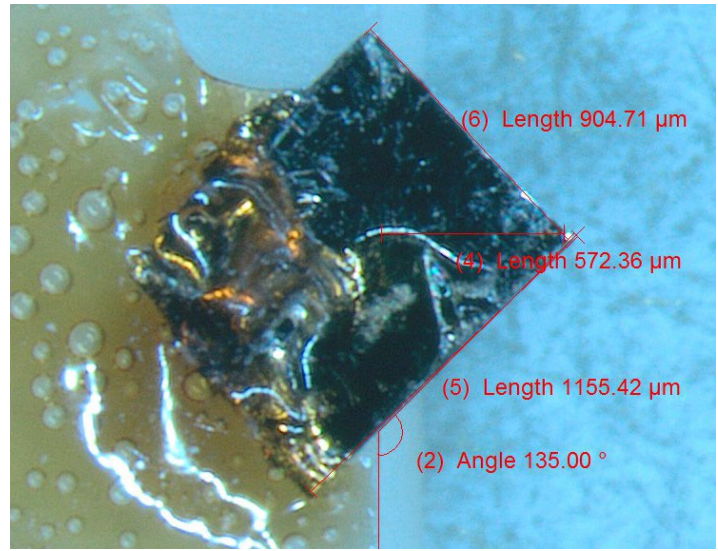

*Fig. 2 Example of a crystal affixed using GE varnish to a sapphire sample mount for measurements at 6-ID-C. The precise alignment angle brings the tetragonal $(220)_T$ Bragg peak in the horizontal scattering plane in-line with the detector.*

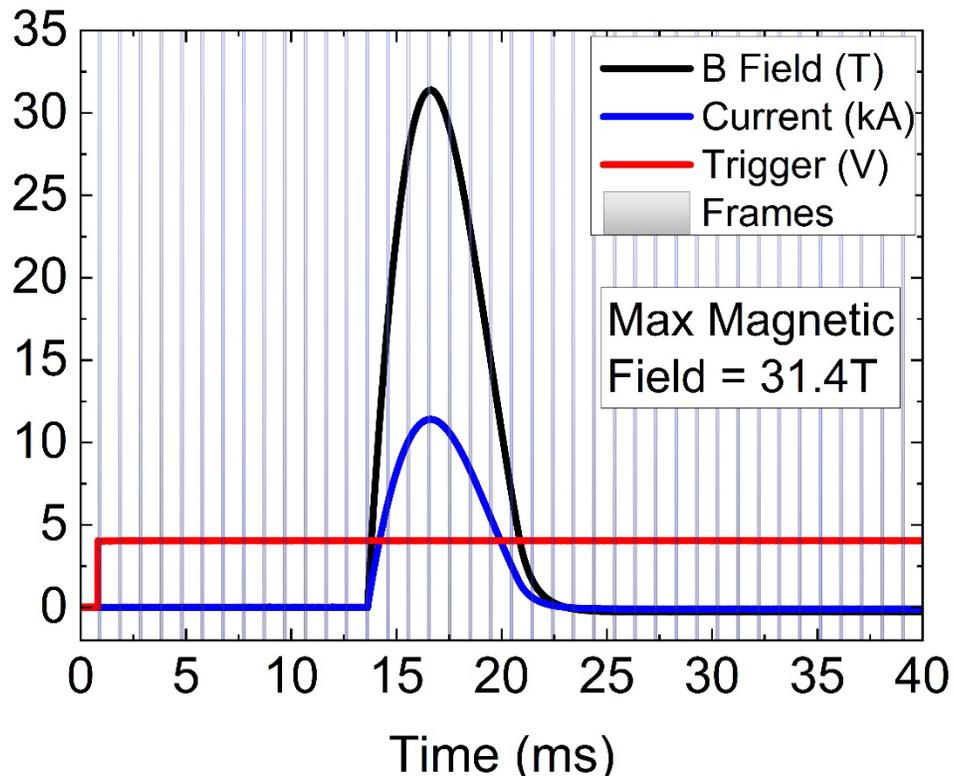

*Fig. 3 Oscilloscope readout during a typical magnetic pulse. The trigger pulse, current, and frame timings were recorded. The trigger pulse initiates the collection of frames with a brief delay to start the pulse so that it reaches its peak on the 17th frame. Some deadtime is included before the magnetic pulse is initiated so that each measurement begins with no field for comparison. The magnetic field, B, was calculated using the coil constant of 2.75 T/kA. The signal begins on the 14th frame and peaks on the 17th.*

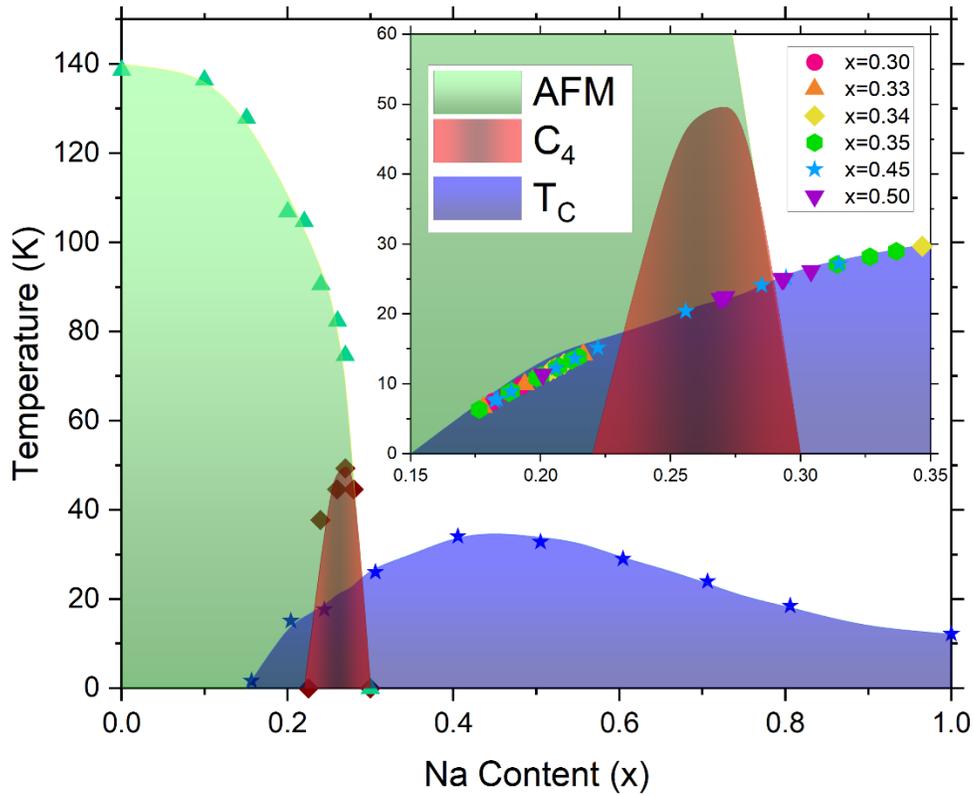

Fig. 4 Phase diagram of $Ba_{1-x}Na_xFe_2As_2$ determined from powder data from refs [3,4]. The inset shows the $T_c$ values of crystals versus effective Na-content obtained according to equation 1.

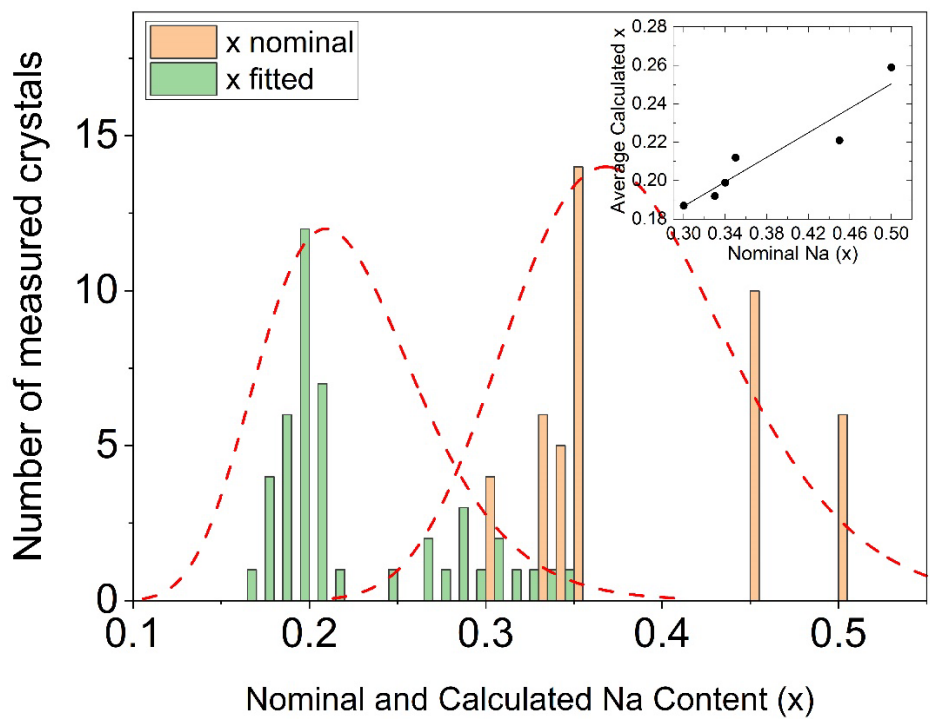

*Fig. 5 Two-dimensional histogram distribution plot comparing the nominal and fitted Na stoichiometries (x) values. The shape of the histograms is an artifact of the number of crystals measured, but the consistent shift demonstrates systematic behavior. Inset: Plot of calculated Na content (averaged over all the best crystals measured from each batch) as a function of nominal stoichiometry. These ratios are valid specifically to the (Ba,Na)- 122 system, but as the chemistry is similar for the rest of the hole-doped 122s, these should be good ballpark starting values to be iterated upon in those related systems.*

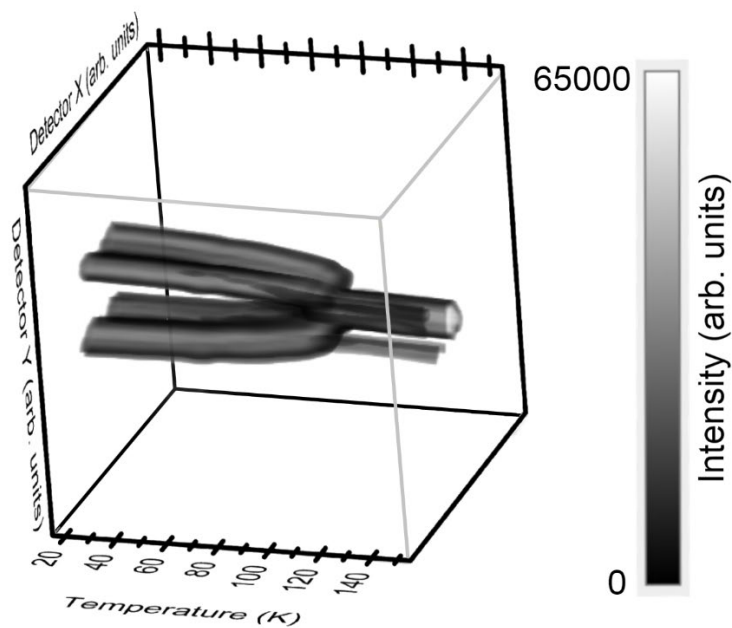

*Fig. 6 Results of temperature-dependent structural measurements on 6-ID-C. The four orthorhombic (200)$_{orth}$ and (020)$_{orth}$ twin-type Bragg peaks merge at T$_S$ into a single (220)$_T$ tetragonal peak. Scans were taken at intervals of 5K between 16 and 140K. ImageJ freeware [74] was used to arrange 2D detector images in stacks and interpolate between to visualize the evolution of the structural transition in 3D. The orthorhombic to tetragonal transition occurs at ~108 K in agreement with the crystal's Na content.*

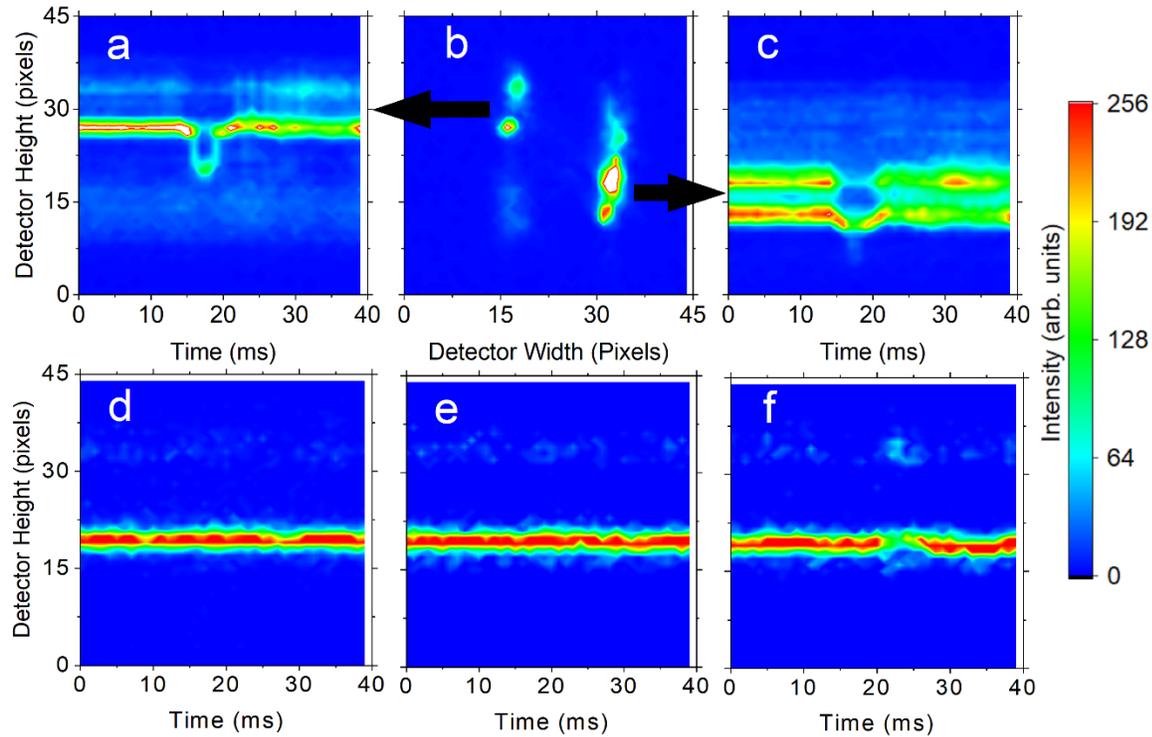

*Fig. 7 (a – c) Effects of a 31.4T magnetic pulse applied along the c-axis on the $(200)_O$ / $(020)_O$ Bragg peaks. The 17th frame corresponds to the maximum intensity of the magnetic pulse, and the effect on the diffraction pattern correlates perfectly with the pulse. The left and right panels (a and c) are vertical slices (in time) through the left and right Bragg peaks in the center image (b), respectively. The data were collected at 95K, just below the orthorhombic transition at ~ 108 K for this sample. No measurements at different temperatures or of different samples showed the same effect. The data used to produce this figure was a summation of the detector counts over 10 identical pulses and measurements, demonstrating the repeatability of the effect. Measurements of the same orthorhombic peaks (only one of the four peaks is shown) performed under magnetic fields of 0.7 (d), 17.5 (e) and 31.4 T (f) at a slightly different incident angle.*

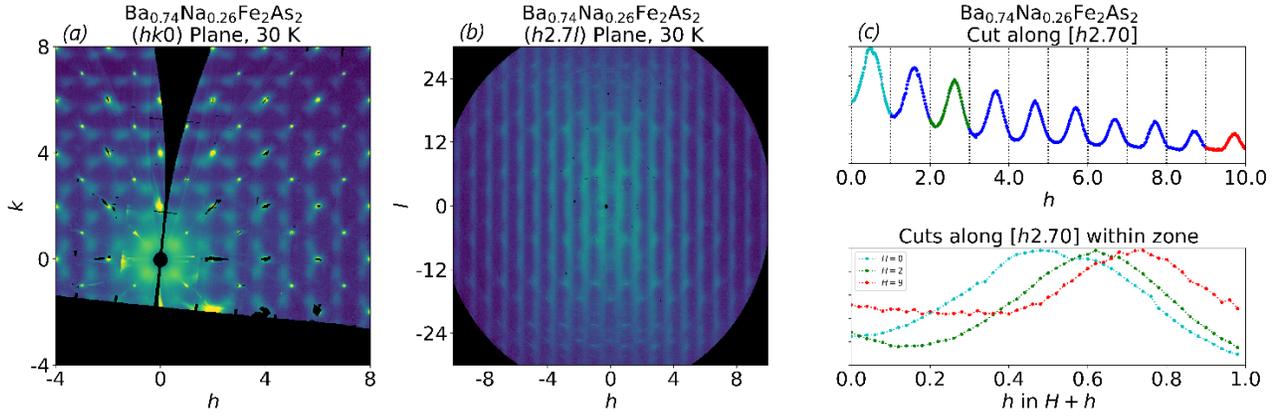

*Fig. 8 Example of single crystal diffuse scattering data obtained at 6-ID-D. The crystal was measured at 30 K and has a superconducting $T_C$ of 22 K, putting it in the middle of the $C^c_{4M}$ dome. (a) shows the (h k 0) plane. (b) shows the (h 2.7 l) plane. (c) shows a broad cut along h, integrated along l = for – 5 to 5 reciprocal lattice units (r.l.u.), with dotted black lines indicating integer values. Cuts starting at H = 0.0, H = 2.0, and H = 9.0 are shown below to emphasize the changing position of this feature.*

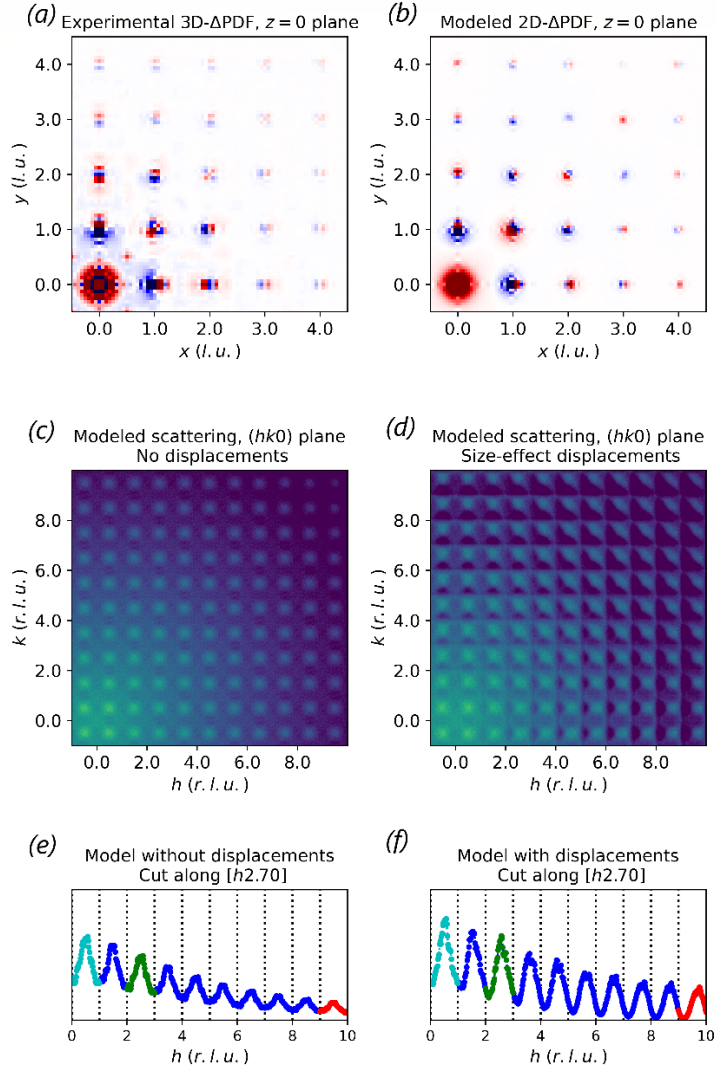

*Fig. 9: Use of 3D-ΔPDF to produce a qualitative model. a) 3D-ΔPDF from diffuse scattering data collected at 30 K. b) 2D-ΔPDF from DISCUS model. This should be comparable to the 3D-ΔPDF data because it does not have any correlations along the c axis. The size effect features are qualitatively reproduced, while subtle displacement correlations (notably at x=2, y=0 and x=0, y=2) are not. c) Modeled scattering intensity prior not including size effect displacement showing broad peaks at h=(n+1/2), k=(n+1/2) positions. d) Modeled scattering intensity including size effect displacements, showing a shift in diffuse scattering similar to that seem in experiment. e) Line cut along [h 2.7 0] from model not including size effect displacements along [h 2.7 0], similar to Fig. 8c. f) Line cut along [h 2.7 0] from model including displacements.*